\documentclass[aps,prb,twocolumn,amsmath,amssymb,nofootinbib,superscriptaddress,floatfix]{revtex4-2}

\usepackage[english]{babel}

\usepackage{color}
\usepackage{xparse}
\usepackage{amsmath}
\usepackage{physics}
\usepackage{braket}
\usepackage{graphicx}
\usepackage{hyperref}
\usepackage{lipsum}
\usepackage{csquotes,helvet,mathpazo,listings}
\usepackage{notes2bib}
\usepackage{dcolumn}
\usepackage{bm}
\usepackage[normalem]{ulem}
\usepackage{verbatim}
\usepackage{pifont}
\newcommand{\bs}[1]{{\boldsymbol{#1}}}

\begin{document}

\title{
{Higher-order Van Hove Singularities in Kagome Topological Bands}
}

\author{Edrick Wang}
\affiliation
{Department  of  Physics,  Emory  University,  400 Dowman Drive, Atlanta,  Georgia  30322,  USA}
\author{Lakshmi Pullasseri}
\affiliation
{Department  of  Physics,  Emory  University,  400 Dowman Drive, Atlanta,  Georgia  30322,  USA}
\author{Luiz H Santos}
\affiliation
{Department  of  Physics,  Emory  University,  400 Dowman Drive, Atlanta,  Georgia  30322,  USA}

\date{\today}

\begin{abstract}
Motivated by the growing interest in band structures featuring higher-order Van Hove singularities (HOVHS), we investigate a spinless fermion kagome system characterized by nearest-neighbor (NN) and next-nearest-neighbor (NNN) hopping amplitudes. While NN hopping preserves time-reversal symmetry, NNN hopping, akin to chiral hopping on the Haldane lattice, breaks time-reversal symmetry and leads to the formation of topological bands with Chern numbers ranging from $C = \pm 1$ to $\pm 4$. 
We perform analytical and numerical analysis of the energy bands near the high-symmetry points $\bs{\Gamma}$, $\pm \bs{K}$, and $\bs{M_i}$ ($i=1,2,$ and $3$), which uncover a rich and complex landscape of HOVHS, controlled by the magnitude and phase of the NNN hopping. We observe power-law divergences in the density of states (DOS), $\rho(\epsilon) \sim |\epsilon|^{-\nu}$, with exponents $\nu = 1/2, 1/3, 1/4$, which can significantly affect the anomalous Hall response at low temperatures when the Fermi level crosses the HOVHS. Additionally, the NNN hopping induces the formation of higher Chern number bands $C = \pm 2, \pm 4$ in the middle of the spectrum obeying a sublattice interference whereupon electronic states are maximally localized in each of the sublattices when the momentum approaches the three high-symmetry points $\bs{M_i}$ ($i=1,2,$ and $3$) on the Brillouin zone boundary. This classification of HOVHS in kagome systems provides a platform to explore unconventional electronic orders induced by electronic correlations.    
\end{abstract}

\maketitle

\section{Introduction}

The interplay between topological 
electronic bands and correlation effects provides a fruitful avenue to realizing unconventional phases driven by competing electronic states. The presence of large density of states (DOS) provides an effective mechanism to enhance electronic correlations. In 2D lattices, Van Hove has shown that saddle points in the band dispersion lead to Van Hove singularities (VHS) characterized by logarithmic divergent DOS \cite{VanHove1953TheOccurrence}. The possibility of higher-order VHS (HOVHS) displaying even stronger power-law divergent DOS, considered in the context of cuprate band structures  \cite{abrikosov1993experimentally,gofron1994observation}, 
has recently garnered significant interest
as a pathway to explore unconventional electronic orders \cite{shtyk2017electrons,wang2021moire,sherkunov2018electronic,yuan2019magic,efremov2019multicritical, 
kerelsky2019maximized,lin2020parquet, 
classen2020competing,hsu2021spin,bi2021excitonic,kang2022twofold,hu2022rich,wu2023pair}.
In particular, when HOVHS emerge in Chern bands 
\cite{castro2023emergence, aksoy2023single, pullasseri2024chern}, the interplay between band topology and high DOS can promote novel electronic orders, 
such as superconducting pair-density waves and Chern supermetals \cite{castro2023emergence}, which underscores the importance of higher-order singularities in topological bands for discovering new quantum phases. However, beyond initial studies on the Haldane lattice  \cite{castro2023emergence, aksoy2023single} and topological insulator moir\'e surface states  \cite{pullasseri2024chern}, the structure of HOVHS in Chern bands remains largely unexplored.

In this work, we extend this knowledge by investigating a rich scenario where band topology and HOVHS are intertwined in a 2D kagome system. Kagome lattices  \cite{yin2022topological, balents2010spin}, characterized by their geometrically frustrated network of corner-sharing triangles in two dimensions, support VHSs \cite{kang2020dirac, hu2022rich}, along with flat bands 
\cite{green2010isolated, ohgushi2000spin, bergman2008band, xiao2003landau, wu2007flat}
and Dirac fermions \cite{ye2018massive, kang2020dirac}, which provide an ideal platform for studying complex electronic phenomena. The substantial enhancement of interaction effects due to the large accumulation of electronic states around the VHS plays a crucial role in the emergence of various electronic phenomena including charge density waves, pair density waves, and unconventional superconductivity \cite{wen2010interaction, wang2013competing, kiesel2013unconventional, lin2021complex,scammell2023chiral, nag2024pomeranchuk,jiang2024van}.

\begin{figure}
    \includegraphics[width= 0.5\linewidth]{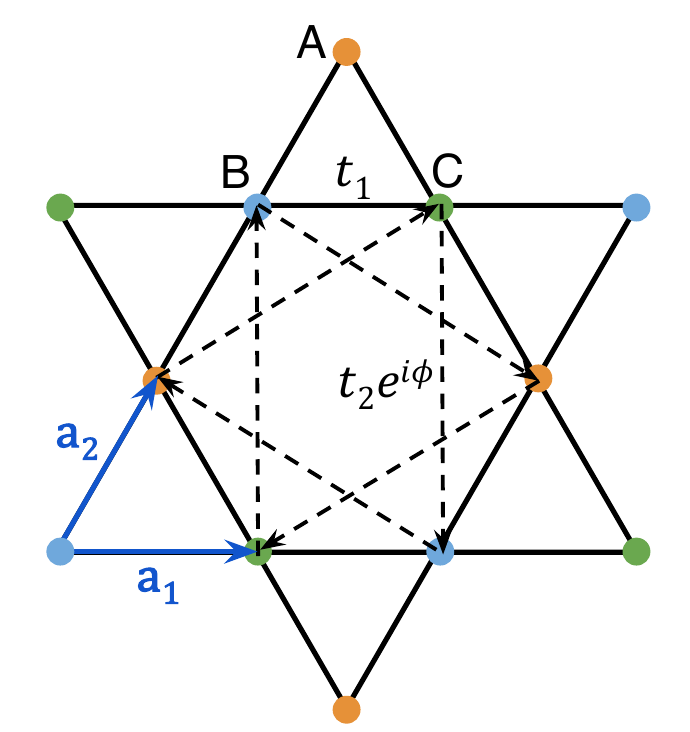}
    \caption{Kagome lattice with the sites A, B, and C marked in orange, blue, and green respectively. The vectors connecting site B with its nearest neighbors A and C are denoted as $\mathbf{a_2}$ and $\mathbf{a_1}$ respectively. The real NN hopping amplitude, $t_1$, is represented by solid black lines. The black dashed arrows show the orientation of the NNN hoppings with strength $t_2 e^{i \phi}$. This NNN hopping breaks time-reversal symmetry while preserving $\mathcal{C}_{3}$ rotation and inversion symmetry about the center of the hexagon.}
    \label{fig: lattice model+band structure}
\end{figure}

We investigate the effects of a complex next-nearest-neighbor (NNN) hopping amplitude, $t_2\,e^{i\phi}$, on the electronic bands of a kagome lattice. This hopping term, analogous to the Haldane model on the honeycomb lattice \cite{Haldane1988}, breaks time-reversal symmetry (TRS) and induces the formation of bands with nonzero Chern numbers. Employing a systematic analytical and numerical classification of critical points at high-symmetry points $\bs{\Gamma}$, $\pm \bs{K}$, and $\bs{M_i}$ ($i = 1, 2, 3$) of the Brillouin zone as a function of $(t_2, \phi)$, we reveal a rich landscape of HOVHS in this time-reversal broken kagome system. Notably, we identify conditions under which bands with Chern numbers as large as $C = \pm 4$, emerge. Owing to the presence of HOVHSs, these bands support power-law divergences in the DOS, $\rho(\epsilon) \sim |\epsilon|^{-\nu}$, with exponents $\nu = \frac{1}{2}$, $\frac{1}{3}$, and $\frac{1}{4}$. 
{In addition to the classification of HOVHS, this TRS-broken system reveals new features of the kagome band structure:}
\\
\noindent
{(1) While the NNN hopping destroys the exact flatness condition of the third band in the nearest-neighbor (NN) kagome lattice \cite{bergman2008band, xiao2003landau, wu2007flat} (a special case where all points in the Brillouin zone become critical at $t_2=0$), our phase diagram uncovers HOVHS lines at the high symmetry points $\bs{K}$ and $\bs{M_i}$ in the $(t_2,\phi)$ parameter space, which converge to a flat band at $(t_2 \rightarrow 0, \phi = \pm \pi/2)$. Thus the kagome lattice provides a relevant setting to study the emergence of HOVHS near stronger DOS singularities due to flatbands.}
\\
\noindent
{(2) In the lowest band, our classification of HOVHS not only identifies the loci of high DOS but also pinpoints the location of a nearly flat Chern band with $C = \pm 1$ in the $(t_2,\phi)$ parameter space. This provides an ideal scenario for exploring competing electronic orders and emergent fractional Chern insulators \cite{Neupert-2011,Sheng-2011,Tang-2011,Sun-2011,Regnault2011} in a partially filled band.}
\\
\noindent
{(3) The NNN chiral hopping, while breaking TRS, preserves the sublattice interference (SI) \cite{kiesel2012sublattice, wu2023sublattice, schwemmer2023pair} of the second band for the entire parameter space $(t_2, \phi)$. SI implies that electronic states associated with $\bs{M_i}$ points at the Brillouin zone boundary are maximally localized on the sublattices A, B, and C, which has a non-trivial effect on interactions when the Fermi energy crosses a HOVHS at the $\bs{M_i}$ points at the zone boundary. Our work thus extends the mechanism of SI in time-reversal broken Chern bands from the Chern number $C = \pm 1$ band on the honeycomb lattice \cite{castro2023emergence} into the realm of topological kagome bands supporting higher Chern numbers, $C = \pm 2$ and $C = \pm 4$.}
\\

This work opens a direction to explore exotic kagome bands in synthetic materials such as optical lattices \cite{PhysRevLett.108.045305}. In particular, an implementation of the complex NNN hopping akin to a Haldane optical lattices \cite{jotzu2014experimental} could be achieved via periodic modulation using piezoelectric actuators.
Moreover, the recent discovery of a new family of kagome metals, AV$_3$Sb$_5$ (A = K, Rb, Cs) displaying a variety of exotic correlated electronic phenomena \cite{kiesel2012sublattice, wu2021nature, jiang2021unconventional, kang2023charge, liang2021three, wang2023anisotropic, li2022rotation} and exhibiting both conventional and higher-order Van Hove singularities \cite{ortiz2019new, ortiz2020cs, jiang2023kagome}, as evidenced by angle-resolved photoemission spectroscopy (ARPES) \cite{kang2022twofold, hu2022rich}, further motivates a deeper exploration of the HOVHS landscape in kagome lattices. 
While kagome systems have been actively studied in connection with Van Hove singularities, the relationship between VHS and non-trivial band topology in these materials remains largely unexplored.

\noindent
\textit{Model --}
We study a tight-binding model of the kagome lattice with lattice constant $a$, consisting of NN hopping as well as complex NNN hopping,
\begin{equation}
    H = -t_1 \sum_{\langle i,j\rangle} c^{\dag}_i c_j -t_2\sum_{\langle \langle i,j \rangle\rangle} e^{i \phi_{ij}} c^{\dag}_i c_j + h.c. ,
    \label{eq:real_space_kagome}
\end{equation}
where $t_1$ ($t_2$) is the NN (NNN) hopping strength, $\phi_{ij}$ is the phase factor associated with the NNN hopping between sites $i$ and $j$, $\langle i,j\rangle$ and $ \langle\langle i,j\rangle\rangle$ indicates the NN and NNN hopping respectively, and $c^{\dag}_{j}$ is the fermionic creation operator at site $j$.  The vectors connecting the NN atomic sites are defined as $\mathbf{a_1}=\frac{a}{2}(1,0)$, $\mathbf{a_2}=\frac{a}{4}(1,\sqrt{3})$ and $\mathbf{a_3}=\mathbf{a_1}-\mathbf{a_2}$, and the NNN hopping vectors are defined as $\mathbf{b_1} = \frac{a}{2} (0, \sqrt{3})$, $\mathbf{b_2} = \frac{a}{4} (3, -\sqrt{3})$, and $\mathbf{b_3} = -(\mathbf{b_1} + \mathbf{b_2})$, as shown in Fig.~\ref{fig: lattice model+band structure}. Notice that the black dashed arrows shown in Fig.~\ref{fig: lattice model+band structure} denote the direction of the complex NNN hopping. Furthermore, we assume the system is spin-polarized, thereby omitting the spin degree of freedom.
\begin{figure}[h!]
    \centering
    \includegraphics[width=0.9\linewidth]{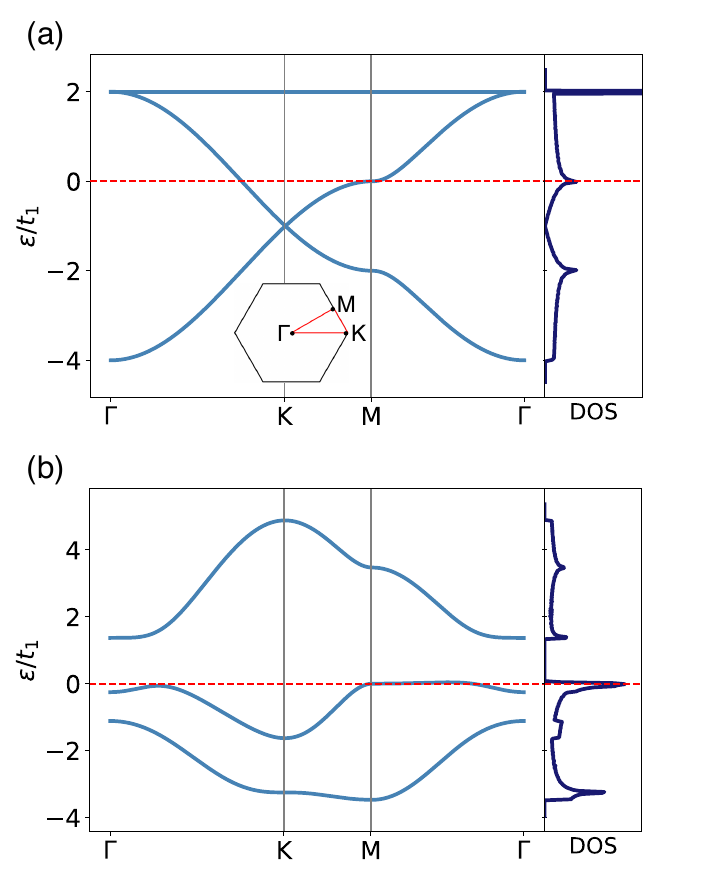}
    \caption{Band structure at (a) $(t_2, \phi) = (0,0)$ and (b) 
    $(t_2, \phi) = (0.76, 0.9 \pi)$, with the first BZ shown in (a). The
    corresponding DOS is displayed on the right. The red dashed lines in (a) and (b) denote the energy at which the $\bs{M}$ points of the second band support a conventional VHS and a HOVHS respectively, as indicated by the divergences in the corresponding DOS. Notice that the energy corresponding to the $\bs{M}$ points on band 2 is always zero.}
    \label{fig: Band structure}
\end{figure}

In the momentum space, the Hamiltonian reads $H=\sum_{\bs{k}}c^{\dag}_{\mathbf{k}} \hat{\mathcal{H}}(\bs{k}) c_{\mathbf{k}}$, where $\hat{\mathcal{H}}(\bs{k})= \hat{\mathcal{H}}_{NN}(\bs{k}) + \hat{\mathcal{H}}_{NNN}(\bs{k})$, and $c_{\mathbf{k}}=(c_{\mathbf{k},\text{A}} \: \ c_{\mathbf{k},\text{B}} \: \ c_{\mathbf{k},\text{C}})^T$, with A, B, C being the three sites of the kagome unit cell displayed in Fig.~\ref{fig: lattice model+band structure}. 
The momentum $\bs{k}$ is defined in the first Brillouin zone (BZ) spanned by the two reciprocal lattice vectors, $\mathbf{G_1}=\frac{2\pi}{a}(1,-\frac{1}{\sqrt{\smash[b]{3}}})$ and $\mathbf{G_2}=\frac{2\pi}{a}(0,\frac{2}{\sqrt{\smash[b]{3}}})$. The lattice constant $a$ will henceforth be set to 1.
The single particle Hamiltonian $\hat{\mathcal{H}}(\bs{k})$ can be expressed as
\begin{equation}
\label{eq: Hamiltonian}
\hat{\mathcal{H}}(\bs{k}) = 
\begin{pmatrix}
0 
&
h_{12}(\bs{k})
&
h_{13}(\bs{k})
\\
h^{*}_{12}(\bs{k})
&
0
&
h_{23}(\bs{k})
\\
h^{*}_{13}(\bs{k})
&
h^{*}_{23}(\bs{k})
&
0
\end{pmatrix}
\,,
\end{equation}
where
$h_{12}(\bs{k}) = 
-2\,t_1\,\cos{(\bs{k}\cdot\bs{a}_{2})} 
-2\,t_2\,e^{i\phi}\,
\cos{(\bs{k}\cdot\bs{b}_{2})} 
$,
$h_{13}(\bs{k}) = 
-2\,t_1\,\cos{(\bs{k}\cdot\bs{a}_{3})} 
-2\,t_2\,e^{-i\phi}\,
\cos{(\bs{k}\cdot\bs{b}_{3})} 
$,
and
$h_{23}(\bs{k}) = 
-2\,t_1\,\cos{(\bs{k}\cdot\bs{a}_{1})} 
-2\,t_2\,e^{i\phi}\,
\cos{(\bs{k}\cdot\bs{b}_{1})} 
$.

Diagonalization of the Hamiltonian given in Eq.~\eqref{eq: Hamiltonian}, $H=\sum_{\bs{k}}\sum_{n=1,2,3}\Psi^{\dag}_{n,\mathbf{k}} \epsilon_{n,\mathbf{k}} \Psi_{n,\mathbf{k}}$, yields the dispersion of each band, $\epsilon_{n,\mathbf{k}}$, where 
$n=1,2,3$ denotes the index of the first, second and third energy bands, respectively, in ascending order. Henceforth, energy is expressed in units of $t_1$.
Owing to $\mathcal{C}_3$ rotation and inversion symmetries, the spectrum satisfies 
$\epsilon_{n,\mathcal{C}_{3}\mathbf{k}}=\epsilon_{n,\mathbf{k}}$
and
$\epsilon_{n,-\mathbf{k}}=\epsilon_{n,\mathbf{k}}$.
\begin{figure}
    \centering
    \includegraphics[width=0.95\linewidth]{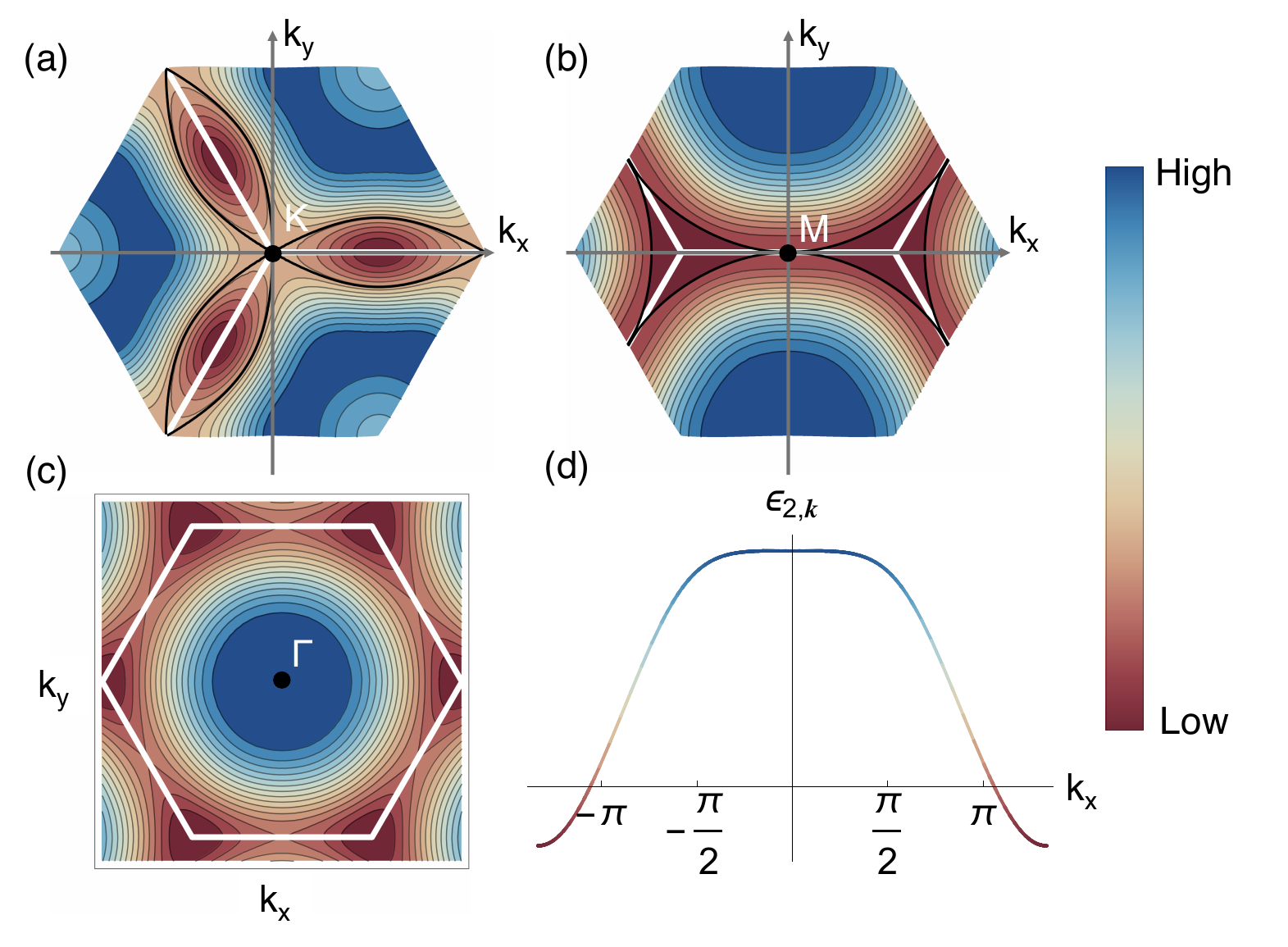}
    \caption{Contour plots of the energy dispersion corresponding to band 2, near the high-symmetry points (a) $\pm\bs{ K}$, (b) $\bs{M_i}$, and (c) $\bs{\Gamma}$ points, where the Hessian vanishes, thereby supporting HOVHS. The white lines denote the boundaries of the first BZ and the black lines correspond to the Fermi surface contour at the corresponding energy of the HOVHS. At the $\bs{K}$ point (a), the dispersion exhibits a monkey-saddle dispersion. In contrast, around the $\bs{M_3}$ point (b), the dispersion becomes locally flat only along the $k_x$ direction. At the $\bs{\Gamma}$ point (c), which is an extremum, the second-order curvature of the dispersion vanishes along both $k_x$ and $k_y$ directions. (d) The energy dispersion considered in (c), around the $\bs{\Gamma}$ point, plotted along $k_x$ with $k_y $ held constant at zero, in order to highlight the flatness of the band around the $\bs{\Gamma}$ point.}
    \label{fig: HOVHS examples}
\end{figure}

\section{Classification of Higher Order Van Hove Singularities} 
\label{sec: HOVHS}

In 2D Bloch bands, an ordinary VHS exhibits logarithmic divergence in the DOS ($\rho \propto \text{log}|\epsilon|$) \cite{VanHove1953TheOccurrence}, which occurs at a saddle point (i.e., where the dispersion is locally $\epsilon_{n, \bs{k}} \sim k^{2}_{x} - k^{2}_{y}$), with the following conditions satisfied: $\nabla_{\mathbf{k}}\epsilon_{n,\mathbf{k}}=0$ and $\mathbb{H}_{n,\mathbf{k}}<0$, where $\mathbb{H}_{n,\mathbf{k}} = \text{det}(\frac{\partial^2\epsilon_{n,\mathbf{k}}}{\partial k_i \partial k_j})$ is the Hessian of the dispersion $\epsilon_{n,\mathbf{k}}$.
{When the Hessian at a critical point vanishes, the quadratic form approaches degeneracy, making the energy dispersion of at least third order. When this happens, HOVHS emerges with a power-law divergence in DOS \cite{shtyk2017electrons,yuan2019magic,chandrasekaran2020catastrophe,wang2021moire} due to a higher-order critical point. To clarify the usage of HOVHS in our paper, we point out that this type of singularity can manifest itself in the form of either a higher-order saddle point or a higher-order extrema. While conventional extrema do not give rise to divergence in DOS, higher-order extrema generate a flat local dispersion, inducing a power-law divergence in DOS, which we will discuss in the following subsections. We also emphasize that in the case of an HOVHS resulting from an extremum, we do not observe the emergence of the singularity from the merging of ordinary VHS on the energy band, which is the case in higher-order saddle points like the monkey saddles \cite{shtyk2017electrons}.}

In the absence of NNN hopping, 
as shown in Fig.~\ref{fig: Band structure} (a), band 3 is flat and bands $1$ and $2$ support critical points on the high-symmetry points
of the first BZ 
$\boldsymbol{\Gamma}=(0,0)$, 
$\mathbf{\pm K}=(\pm\frac{4\pi}{3},0)$, 
$\mathbf{M_1}=(\pi,\frac{\pi}{\sqrt{3}})$, 
$\mathbf{M_2}=(-\pi,\frac{\pi}{\sqrt{3}})$, 
and $\mathbf{M_3}=(0,-\frac{2\pi}{\sqrt{3}})$, with conventional saddle points located on $\bs{M_i}$. 
Figure  ~\ref{fig: Band structure} (b) illustrates the effect of the NNN hopping on the band structure, where we observe the onset of bands with significantly higher DOS divergence than ordinary VHS. Henceforth we focus on addressing how these critical points emerge as a function of the parameters $(t_2, \phi)$.
As shown below, all high-symmetry points on the three bands can support HOVHS in this parameter space. In particular, we analyze high-symmetry points separately and classify their HOVHS as shown in Fig.~\ref{fig: HOVHS examples}. $\mathcal{C}_3$ rotation and inversion symmetries, reduce the analysis to three of the six high-symmetry points.

\subsection{Critical points at $\mathbf{\pm K}$} 
Higher-order singularities emerge at $\bs{K}$ in the form of monkey saddles, as shown in Fig.~\ref{fig: HOVHS examples} (a). The corresponding low-energy dispersion, which reflects the $\mathcal{C}_3$ rotation symmetry around the $\bs{K}$ points, takes the form
\begin{equation}
     \epsilon_{\mathbf{K} + \mathbf{p}} - \epsilon_{\mathbf{K}} =  \alpha (p_x^3-3p_xp_y^2)  + \cdot\cdot\cdot,
\end{equation}
where $p_x$ and $p_y$ are defined parallel and perpendicular to the $\bs{\Gamma K}$ line in the first Brillouin zone, respectively,
and the coefficient $\alpha$ is real. 
This monkey saddle dispersion indicates that the corresponding DOS exhibits a power-law divergence with exponent $\nu = 1/3$ \cite{wang2021moire, chandrasekaran2020catastrophe, PhysRevB.101.125120, shtyk2017electrons}, which we confirm numerically. Furthermore, these observations apply to the $-\bs{K}$ points as well, since the model is symmetric under inversion.
\begin{figure*}[t]
    \includegraphics[width=0.9\linewidth]{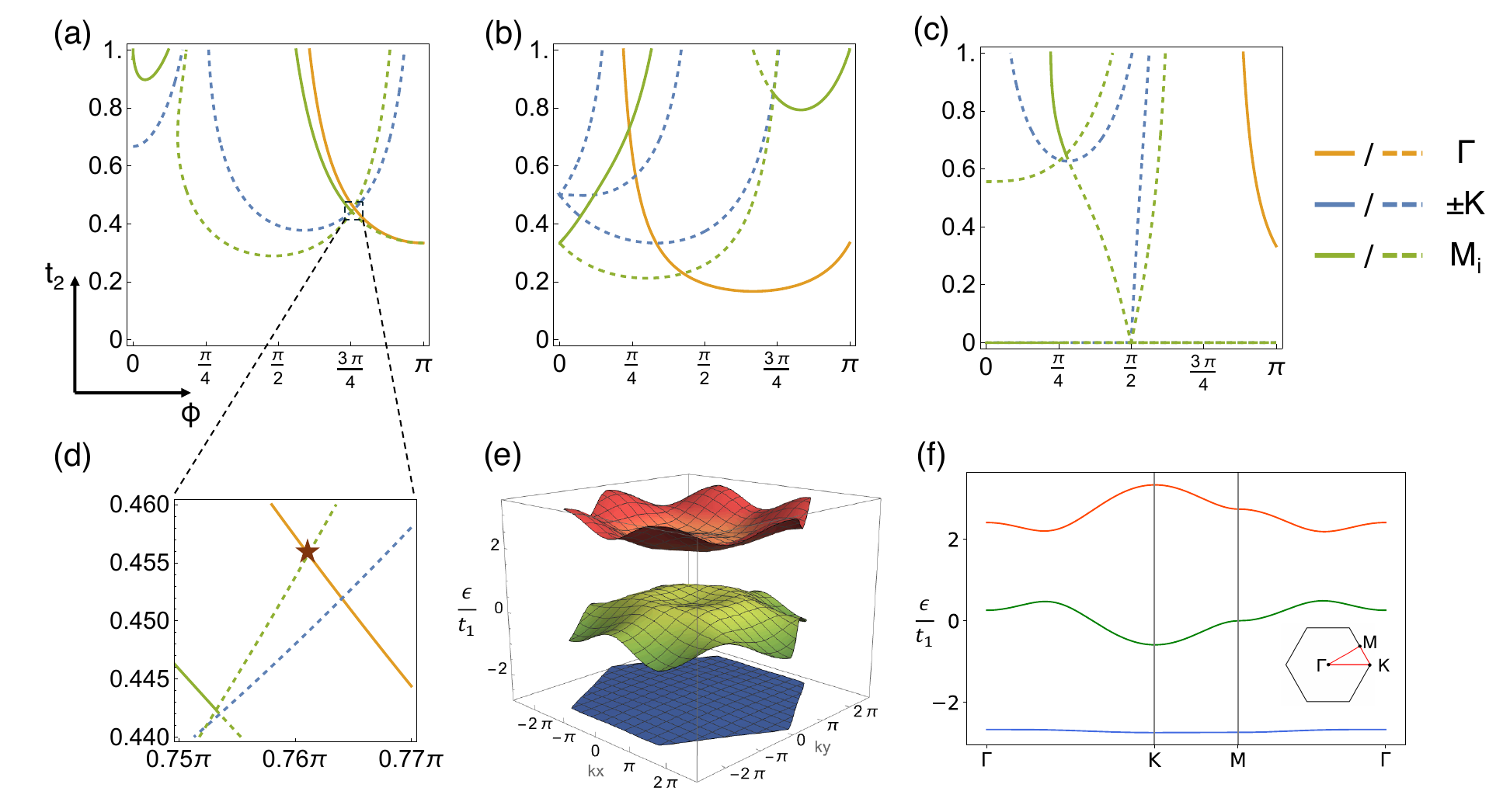}
    \caption{(a)-(c) Parameter space plots showing the set of ($t_2, \phi$) values for which the high-symmetry points $\bs{\Gamma}$(orange), $\bs{\pm K}$ (blue) and $\bs{M}$({green}) corresponding to the bands $1$-$3$ (from left to right) support HOVHSs. The dashed lines correspond to the high-symmetry points which can be classified as higher-order saddles whereas the solid lines correspond to extrema with vanishing Hessian. (d) Parameter space plot zoomed in around the intersections of the lines highlighted with the black dashed lines in (a), clarifying the different intersections of boundary lines. The parameter pair of interest is labeled with $\star$, with values $(t_2=0.45,\phi=0.76\pi)$. (e) The 3D plot of the band structure at parameters labeled with $\star$ in (d), supporting HOVHSs at both $\bs{M}$ and $\bs{\Gamma}$ points. {The energy value corresponding to the HOVHS at the $\bs{M}$ and $\bs{\Gamma}$ points are $-2.74 \, t_1$ and $-2.67 \, t_1$ respectively.} (f) Band diagram of the same bands to showcase the exceptional flatness of band 1, with an approximate bandwidth of $0.08\,t_1$. The first BZ is shown in the diagram as well. 
    }
    \label{fig: HOVHS Phase Diagram+Intersection}
\end{figure*}
\subsection{Critical points at \texorpdfstring{$\boldsymbol{\Gamma}$}{Gamma}}

In contrast to the $\bs{\pm K}$ points, the $\bs{\Gamma}$ point does not support a high-order saddle even though the Hessian $\mathbb{H}_{n,\bs{\Gamma}}$ vanishes. Instead, as the second-order curvature of the dispersion vanishes along both the $k_x$ and $k_y$ directions, $\bs{\Gamma}$ becomes an extremum with vanishing Hessian, thereby resulting in a locally flat band around the $\bs{\Gamma}$ point, as shown in Fig.~\ref{fig: HOVHS examples} (c). The corresponding low-energy dispersion takes the form,
\begin{equation}
\label{eq: gamma dispersion quartic}
\epsilon_{\boldsymbol{\Gamma} + \mathbf{p}} - \epsilon_{\boldsymbol{\Gamma}} = \alpha(p^{2}_{x}+p^{2}_{y})^{2} = \alpha\,p^{4}, 
\end{equation}
where $\alpha$ is a real parameter and $(p_{x},p_{y}) = p\,(\cos{\theta},\sin{\theta})$. Furthermore, the dispersion exhibits a stronger power-law divergence with exponent $\nu = 1/2$, i.e., $\rho(\epsilon) \propto |\epsilon - \epsilon_{\bs{\Gamma}}|^{-\frac{1}{2}}$.
Additionally, we notice one particular instance at $(t_2 \,, \phi) = (1/3 \,, \pi)$ for the first band where the coefficient $\alpha$ vanishes resulting in a low-energy dispersion of the form (up to $\mathcal{O}(p^6)$)
\begin{equation}
\label{eq: epsilon gamma 2}
\epsilon_{\boldsymbol{\Gamma} + \mathbf{p}} 
-
\epsilon_{\bs{\Gamma}}
\approx  \frac{p_x^6}{1152}- \frac{p_x^4 p_y^2}{192} + \frac{p_x^2 p_y^4}{128}
=
\frac{p^6}{1152}\,\cos^{2}{(3\,\theta)}
\,,
\end{equation}
where $\epsilon_{\bs{\Gamma}} = -\frac{8}{3} $, and the corresponding DOS diverges around $\bs{\Gamma}$ as $\rho (\epsilon) \sim |\epsilon - \epsilon_{\bs{\Gamma}}|^{-2/3}$.
We note that both expressions Eqs.~\eqref{eq: gamma dispersion quartic} and \eqref{eq: epsilon gamma 2}
obey inversion and $\mathcal{C}_3$ rotation symmetries.

\begin{figure*}[htbp]
    \includegraphics[width=0.9\linewidth]{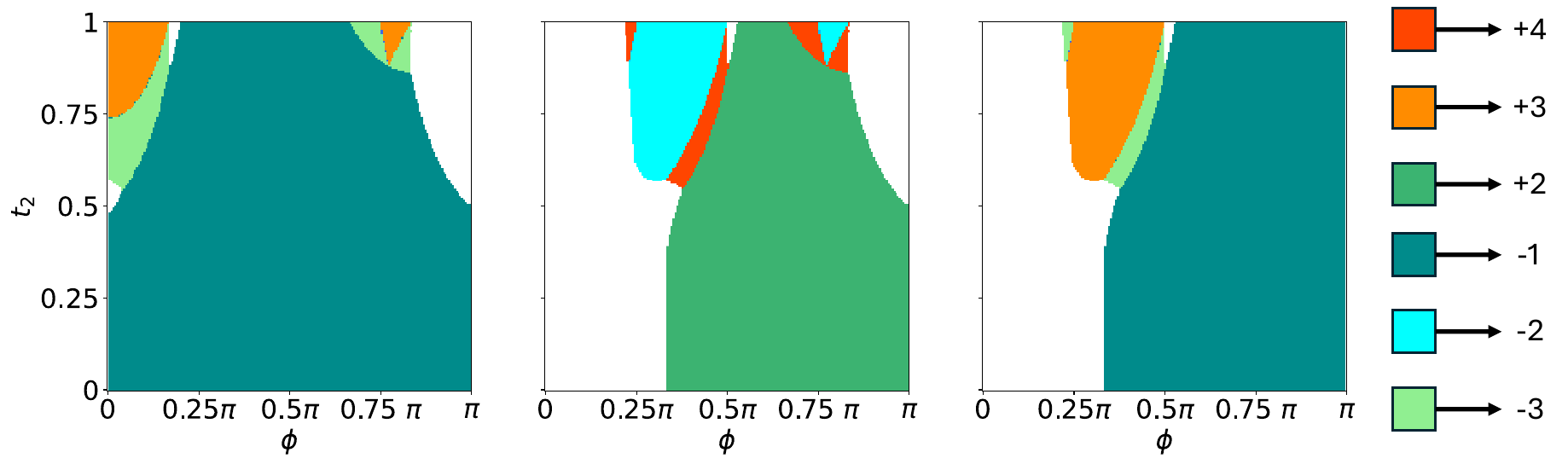}
    \caption{Phase diagrams for band 1 (left), band 2 (middle), and band 3 (right), showing the corresponding Chern numbers in the $t_2$-$\phi$ parameter space. {Since TRS is respected at $\phi=0, \pi$, the Chern number for all three bands at $\phi = 0, \pi$ are $0$.} White regions in the phase diagrams indicate non-positive indirect energy gaps where the Chern number is not well-defined. {(In this case, there is no situation where the Fermi energy lies in between the bands, and the bands are characterized by a possible non-quantized anomalous Hall response.) 
}
    Only $[0,\pi]$ is shown on the $\phi$ axis, since for any Chern number $C$ at $(t_2,\phi)$, as $\phi$ goes to $-\phi$, the Chern number flips sign. As seen in certain regions of the parameter space, band 2 can support Chern number as high as $\pm 4$. }
    \label{fig: Chern number}
\end{figure*}

\subsection{Critical points at $\mathbf{M_i}$}

The emergence of HOVHS at $\bs{M_i}$ points occurs under two conditions: when either one or both of the eigenvalues of the Hessian $\mathbb{H}_{n,\bs{M_i}}$ vanish. Given the model has $\mathcal{C}_3$ symmetry, the following discussion applies to all the $\bs{M_i}$ points, and hence we drop the subscript $i$ for the $\bs{M_i}$ points. When one eigenvalue vanishes, the second-order curvature of the dispersion vanishes along either the $k_x$ or $k_y$ direction, resulting in a locally flat band structure in that direction, as shown in 
Fig.~\ref{fig: HOVHS examples} (b). 
The corresponding low-energy dispersion takes the form,
\begin{equation}
    \begin{split}
    & \epsilon_{\mathbf{M} + \mathbf{p}} - \epsilon_{\mathbf{M}} = \\
    &\begin{cases}
        \alpha p_x^2 + \beta p_x^4 + \gamma p_x^2 p_y^2 + \delta p_y^4 + \cdot\cdot\cdot, 
        \quad 
         \partial_{ky}^2\epsilon_{n,\mathbf{M}}= 0\\
        \alpha p_y^2 + \beta p_x^4 + \gamma p_x^2 p_y^2 + \delta p_x^4 + \cdot\cdot\cdot, 
        \quad 
         \partial_{kx}^2\epsilon_{n,\mathbf{M}}= 0
    \end{cases},
    \label{eq: M points dispersion relation}
    \end{split}
\end{equation}
with all coefficients being real, where the DOS diverges around $\bs{M}$ as $\rho (\epsilon) \sim |\epsilon - \epsilon_{\bs{M}}|^{-1/4}$. On the other hand, when both eigenvalues vanish, the $\alpha$ coefficient vanishes, leading to a fourth-order dispersion in momentum, where the DOS obeys a power-law divergence with exponent $\nu = 1/2$, similar to the case of the $\bs{\Gamma}$ point. { Here, the second-order curvature vanishes in all directions, making the dispersion at $\bs{M_i}$ locally flat.}

\subsection{HOVHS Phase Diagrams}

We numerically calculate the Hessian for the band dispersions corresponding to the Hamiltonian in Eq.~\eqref{eq: Hamiltonian} around the high-symmetry points $\pm\bs{K}$, $\bs{\Gamma}$, and $\bs{M}$. Our analysis reveals a range of $(t_2, \, \phi)$ values, shown in Fig.~\ref{fig: HOVHS Phase Diagram+Intersection} (a)-(c), for which all three bands support HOVHS at one or more of these high-symmetry points. The critical points marked by the HOVHS lines in Figs.~\ref{fig: HOVHS Phase Diagram+Intersection} (a)-(c) all correspond to a vanishing Hessian. Dashed lines indicate higher-order saddle points, while solid lines denote higher-order extrema. 
Notice that the higher-order critical point at $\bs{\Gamma}$ is invariably an extremum, whereas those at $\pm\bs{K}$ are consistently higher-order saddles. Additionally, we find that the Hessian is invariant under the transformation $\phi \rightarrow -\phi$. As a result, we display the HOVHS lines for only $0 \leq \phi \leq \pi$.

{We wish to point out an interesting feature of the flat band in Fig.~\ref{fig: HOVHS Phase Diagram+Intersection} (c). As $t_2$ approaches $0$ around $\phi = \pm \pi/2$, two HOVHS lines of $\bs{M_i}$ and one of $\bs{\pm K}$ are seen merging at a single parameter value. This convergence suggests that, as we perturb the system away from $t_2=0$ around $\phi=\pm \pi/2$, even when the global band dispersion is no longer perfectly flat, the local energy dispersion around the $\bs{\pm K}$ and $\bs{M_i}$ points remains relatively flat. This provides a unique setting to investigate the emergence of HOVHS near stronger DOS singularities due to flat bands.}

{Additionally}, we notice a region in Fig.~\ref{fig: HOVHS Phase Diagram+Intersection} (a) where numerous line intersections occur, for the first band. That region is zoomed in and shown in Fig.~\ref{fig: HOVHS Phase Diagram+Intersection} (d). Coincidentally, the bandwidth in that region also achieves a minimum as low as $\approx 0.08 \, t_1$. 
Selecting a parameter pair of $(t_2=0.45,\phi=0.76\pi)$, which is marked with $\star$ in Fig.~\ref{fig: HOVHS Phase Diagram+Intersection} (d), we obtain a band 1 giving rise to HOVHSs both at $\bs{\Gamma}$ and $\bs{M}$ points. The 3D plot of the band structure is showcased in Fig.~\ref{fig: HOVHS Phase Diagram+Intersection} (e), and the 1D band structure of the system is demonstrated in Fig.~\ref{fig: HOVHS Phase Diagram+Intersection} (f), exhibiting the extreme flatness of the lowest band. The particular region shown in Fig.~\ref{fig: HOVHS Phase Diagram+Intersection} (d) can be a promising territory to observe strongly correlated phenomena in the system.

We now discuss an analytic approach underlying the phase diagrams shown in Figs.~\ref{fig: HOVHS Phase Diagram+Intersection} (a)-(c). In particular, we perturbatively determine the quadratic form near each high-symmetry point. For concreteness, we focus on the $\bs{\Gamma}$ point, where nondegenerate perturbation theory holds, except when degeneracies occur between bands 2 and  3 at $\phi=0, \pi$. Expanding the Hamiltonian about $\bs{\Gamma}$, we get

\begin{equation}
\begin{split}
    \hat{\mathcal{H}}(\bs{\Gamma}+\bs{p})&  =  \hat{\mathcal{H}}(\bs{\Gamma}) + \hat{\mathcal{H}}(\bs{p})
\end{split}
\,,
\end{equation}
\begin{equation}
    \hat{\mathcal{H}}(\bs{p})=
\begin{pmatrix}
    0 & \zeta_{12}(\bs{p}) & \zeta_{13}(\bs{p})\\
    \zeta_{12}^*(\bs{p}) & 0 & \zeta_{23}(\bs{p})\\
   \zeta_{13}^*(\bs{p}) & \zeta_{23}^*(\bs{p}) & 0
\end{pmatrix}
\,,
\end{equation}
where, up to second order in momentum,
$\zeta_{12} = t_1(\bs{p} \cdot \bs{a_2})^2+t_2e^{i\phi}(\bs{p} \cdot \bs{b_2})^2$,
$\zeta_{13} = t_1(\bs{p} \cdot \bs{a_3})^2+t_2e^{-i\phi}(\bs{p} \cdot \bs{b_3})^2$,
$\zeta_{23} = t_1(\bs{p} \cdot \bs{a_1})^2+t_2e^{i\phi}(\bs{p} \cdot \bs{b_1})^2$.
The eigenstates of $\hat{\mathcal{H}}(\bs{\Gamma})$ can be expressed as,
$\Psi_{1,\bs{\Gamma}}^{(0)}=\frac{1}{\sqrt{3}}\left(1,1,1\right)$,
$\Psi_{2,\bs{\Gamma}}^{(0)}= \left(\frac{1}{6} \left(-\sqrt{3}-3 i\right),\frac{1}{6} \left(-\sqrt{3}+3 i\right),\frac{1}{\sqrt{3}}\right)$, and
$\Psi_{3,\bs{\Gamma}}^{(0)}=\left(\frac{1}{6} \left(-\sqrt{3}+3 i\right),\frac{1}{6} \left(-\sqrt{3}-3 i\right),\frac{1}{\sqrt{3}} \right)$.

The dispersion, to leading quadratic order, follows from
\begin{equation}
    \epsilon_{n,\bs{\Gamma}+\bs{p}} = \epsilon_{n,\bs{\Gamma}}^{(0)} + \Psi_{n,\bs{\Gamma}}^{(0)*}\hat{\mathcal{H}}(\bs{p})\Psi_{n,\bs{\Gamma}}^{(0)}.
    \label{eq: PT energy}
\end{equation}
and, hence, the Hessian $\mathbb{H}_{n,\bs{\Gamma}}(t_2,\phi)$ of each band:
\begin{equation}
\begin{split}
&\,
\mathbb{H}_{1,\bs{\Gamma}} = \left[ \frac{3}{4}\left(t_1+3t_2\cos(\phi)\right)\right]^2
\,,
\\
&\,\mathbb{H}_{2,\bs{\Gamma}} = \left[-\frac{1}{8}t_1\left(1+3t_2\cos(\phi)-3\sqrt{3}t_2\sin(\phi)\right)\right]^2
\,,
\\
&\,
\mathbb{H}_{3,\bs{\Gamma}} = \left[-\frac{1}{8}t_1\left(1+3t_2\cos(\phi)+3\sqrt{3}t_2\sin(\phi)\right)\right]^2
\,.
\end{split}    
\end{equation}
Setting $\mathbb{H}_{n,\bs{\Gamma}}=0$, results in the HOVHS contours at $\bs{\Gamma}$ shown in Fig.~\ref{fig: HOVHS Phase Diagram+Intersection} (orange lines),
\begin{equation}
\label{eq: tau Gamma}
\begin{split}
&\,
\tau_{1,\bs{\Gamma}}(\phi)=-\frac{\sec(\phi)}{3}
\,,
\\
&\,
\tau_{2,\bs{\Gamma}}(\phi)=-\frac{1}{3(\cos(\phi)-\sqrt{3}\sin(\phi))}
\,,
\\
&\,
\tau_{3,\bs{\Gamma}}(\phi)=-\frac{1}{3(\cos(\phi)+\sqrt{3}\sin(\phi))}
\,.
\end{split}    
\end{equation}

The other high-symmetry points can be dealt with similarly. In particular, the analysis of degenerate perturbation theory for $\bs{\Gamma}$ at $\phi=\pi$, as well as the Hessian expressions for $\bs{M}$ and $\bs{K}$ points are provided in Appendix.~\ref{appendix: Perturbation Theory}.

\begin{figure*}[t]
    \includegraphics[width=\linewidth]{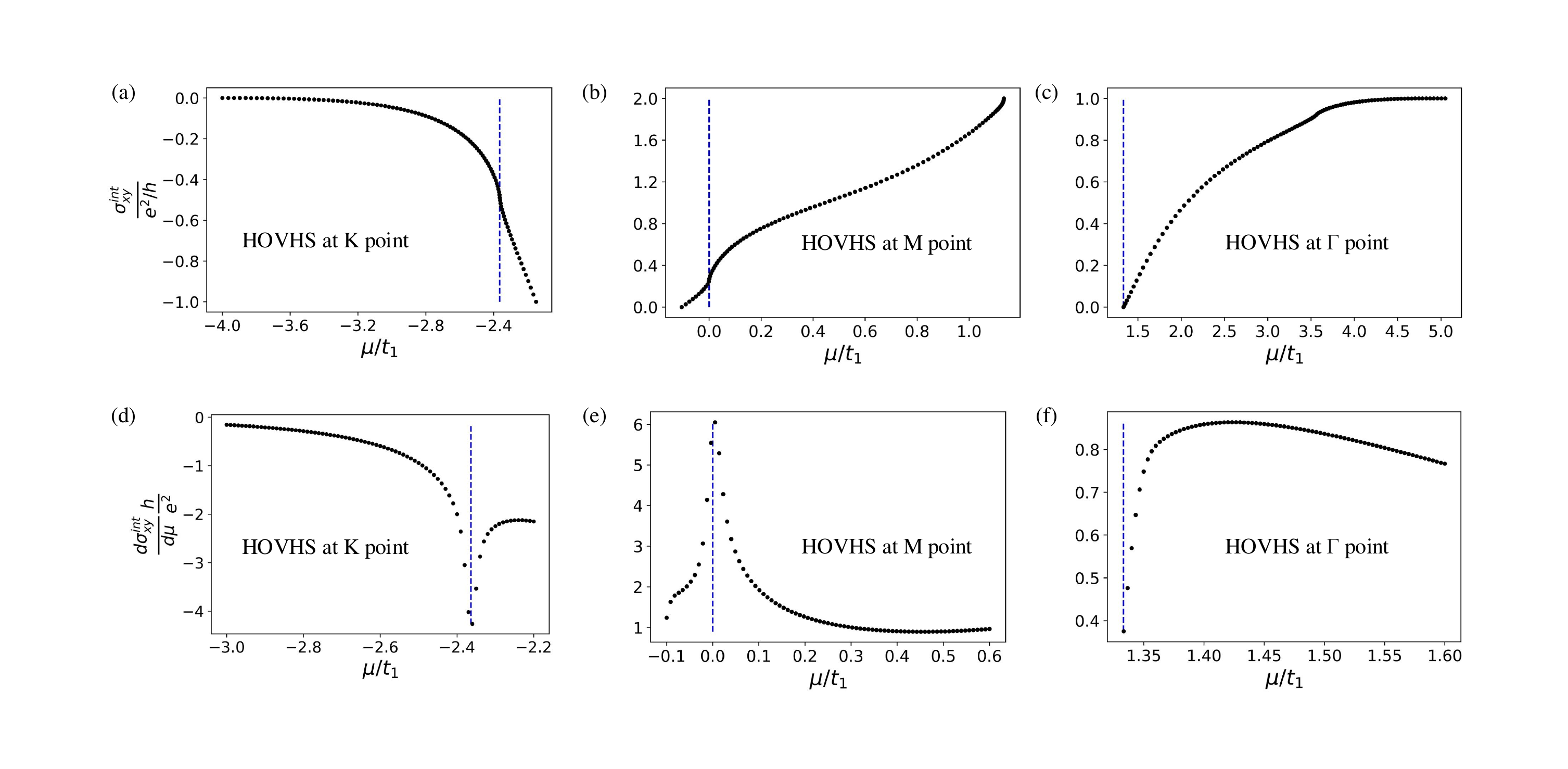}
    \caption{(a-c) Intrinsic anomalous Hall conductivity $\sigma^{\textrm{int}}_{xy}(\mu;0)$ and (d-f) the corresponding differential anomalous Hall conductivity  $\frac{ d \sigma^{\textrm{int}}_{xy}(\mu;0)}{ d \mu}$ at zero temperature, plotted as a function of the Fermi energy $\mu$ in units of $t_1$ for the Chern bands supporting HOVHS at $(t_2,\phi) = (0.39, \pi/2)$ for band 1 (left), $(t_2,\phi) = (0.26, \pi/2)$ for band 2 (middle) and $(t_2,\phi) = (0.80, -0.9\pi)$ for band 3 (right). The high-symmetry point where the HOVHS is located is mentioned in each plot. The differential anomalous Hall response, like the corresponding DOS, exhibits a power-law divergence around $\mu^{*}$ (marked by the blue dashed line) with exponents $1/3$, $1/4$ and $1/2$ for $\bs{K}$, $\bs{M}$ and $\bs{\Gamma}$ point respectively.}
    \label{fig: SigmaAH}
\end{figure*}

\section{Band Topology} \label{subsec: BandTopology}

The NNN hopping breaks time-reversal symmetry and leads to the possibility of energy-isolated topological bands characterized by a nonzero Chern number. To investigate that, we numerically compute \cite{fukui2005chern} the Chern number of the three bands in the $t_2 - \phi$ parameter space.

The Chern number diagrams are presented in Fig.~\ref{fig: Chern number}. We restrict our analysis to $0 \leq t_2 \leq 1$ where the magnitude of the second neighbor hopping is less than the NN hopping $t_1$. Furthermore, the parameter $\phi$ is restricted from $[0,\pi]$, since for any Chern number $C$ at $(t_2,\phi)$, as $\phi$ goes to $-\phi$, the Chern number flips sign. 
The white regions describe bands that are not separated by a direct gap.
The colored regions represent isolated bands with nonzero Chern numbers. In this parameter regime, we observe a rich landscape of gapped topological bands, some of which support relatively large Chern numbers.

By combining the phase diagrams shown in Fig.~\ref{fig: HOVHS Phase Diagram+Intersection} and Fig.~\ref{fig: Chern number}, we uncover a comprehensive understanding of how the NNN hopping leads to the onset of topological bands supporting HOVHS at the high-symmetry points, characterized by power-law diverging DOS $\rho(\epsilon) \sim |\epsilon-\epsilon_{0}|^{-\nu}$, with exponents $\nu = 1/2, 1/4, 1/3$, which is one of the main results of our analysis.

This structure of HOVHS can be accessed upon changing the Fermi energy in each of the bands, which also changes the anomalous Hall response owing to the presence of a finite Berry curvature in the bands. Importantly, the zero temperature differential anomalous Hall response displays a strong divergence whenever the Fermi level crosses a Van Hove singularity, due to the large change in the number of electronic states in a small energy window. At zero-temperature, this response, $\frac{ d \sigma^{\textrm{int}}_{xy}(\mu; T=0)}{ d \mu}$, near VHSs can be expressed in terms of the density of states $\rho(\mu)$ as \cite{pullasseri2024chern} 

\begin{equation}    
\label{eq: general relation sigma and DOS}
\frac{ d \sigma^{\textrm{int}}_{xy}(\mu;T=0)}{ d \mu}= \frac{e^2} {2 \pi h}\, \langle \Omega \rangle_{FS} \, \rho(\mu)
\,,
\end{equation}
where $\sigma^{\textrm{int}}_{xy}(\mu; T=0) = \frac{e^2} {h}\, \frac{1}{2 \pi}\,\int\,d^{2}\bs{k}\,  \Omega(\bs{k}) \Theta(\mu-\epsilon(\bs{k}))$ is the anomalous Hall conductivity at zero-temperature, defined in terms of the Berry curvature $\Omega(\bs{k})$, and
$\langle \Omega \rangle_{FS}$ defines the average of the Berry curvature on the Fermi surface.

In Fig.~\ref{fig: SigmaAH}, we plot $\sigma^{\textrm{int}}_{xy}(\mu;0)$ as well as $\frac{ d \sigma^{\textrm{int}}_{xy}(\mu;0)}{ d \mu}$ as a function of the Fermi energy $\mu$ for three different cases where a Chern band supports HOVHS at either of the three high-symmetry points. Notice that in all the three cases, the inverse of the slope of the anomalous Hall response $\sigma^{\textrm{int}}_{xy}(\mu;0)$ vanishes as $\mu \rightarrow \mu^{*}$ where $\mu^{*}$ is the energy corresponding to the HOVHS, shown in Figs.~\ref{fig: SigmaAH} (a)-(c). As a result, the $\frac{ d \sigma^{\textrm{int}}_{xy}(\mu;0)}{ d \mu}$ plots show a divergence as $\mu \rightarrow \mu^{*}$, displayed in Figs.~\ref{fig: SigmaAH} (d)-(f). 

We numerically confirm that the differential anomalous Hall responses exhibit power-law divergences with the same exponent as the corresponding DOS, and note that the distinctly sharp peaks become progressively broadened as temperature increases.

\section{Sublattice Interference} \label{subsec: SI}

A remarkable feature of Kagome systems is that Bloch states of the second band corresponding to each $\bs{M_i}$ point in the Brillouin zone are maximally localized on one of the three sublattices A, B and C. Notably, this form of sublattice interference (SI) 
\cite{kiesel2012sublattice,wu2023sublattice,kiesel2013unconventional}
is persistent even with long-range real hopping terms extending up to the third nearest neighbor \cite{wu2023sublattice}. However, SI in kagome systems supporting topological bands is yet to be explored. In this section, we investigate
the rich interplay between SI and HOVHS promoted by the complex hopping.

First, we establish that the SI \textit{persists on band 2 throughout the $t_2$-$\phi$ parameter space}.
For instance, consider the high-symmetry point $\bs{M_1} = (\pi \,, \frac{\pi}{\sqrt{3}})$. Working out the dot product of $\bs{M_1}$ with the lattice vectors $\bs{a_i}$ and $\bs{b_i}$, $\bs{M_1} \cdot \bs{a_1} = \pi/2$, $\bs{M_1} \cdot \bs{a_2} = \pi/2$, $\bs{M_1} \cdot \bs{a_3} = 0$, $\bs{M_1} \cdot \bs{b_1} = \pi/2$, $\bs{M_1} \cdot \bs{b_2} = \pi/2$ and $\bs{M_1} \cdot \bs{b_3} = -\pi$, yields the Hamiltonian $\hat{\mathcal{H}}(\bs{M_1})$
\begin{equation}
    \hat{\mathcal{H}}(\bs{M_1}) = 
    -2\,t_1\,
    \begin{pmatrix}
    0 & 0 & 1
    \\
    0 & 0 & 0
    \\
    1 & 0 & 0
    \end{pmatrix}
    -2\,t_2\,
    \begin{pmatrix}
    0 & 0 & -e^{i\phi}
    \\
    0 & 0 & 0
    \\
    -e^{-i\phi} & 0 & 0
    \end{pmatrix}
    \,,
    \label{eq: M Hamiltonian}
\end{equation}
supporting the energy values, $\epsilon_{1,3} = \pm 2 \sqrt{(t_1 - t_2 e^{-i \phi})(t_1 - t_2 e^{i \phi})}$
and $\epsilon_2 = 0$, where $\epsilon_{1} \leq \epsilon_{2} \leq \epsilon_{3}$. The corresponding eigenstates are denoted as $\Psi_{n,\mathbf{k}}=u_{\lambda,\mathbf{k}}c_{\mathbf{k}}$, where $u_{\lambda,\mathbf{k}}$ is a unitary transformation with $\lambda=$ A, B, C. Notably, the $\epsilon = 0$ case corresponds to the second band with eigenvector $\Psi_{2,\bs{M_1}}=(0 \: 1 \: 0)^T$, localized on the B sublattice.

\begin{figure}[h]
    \includegraphics[width=\linewidth]{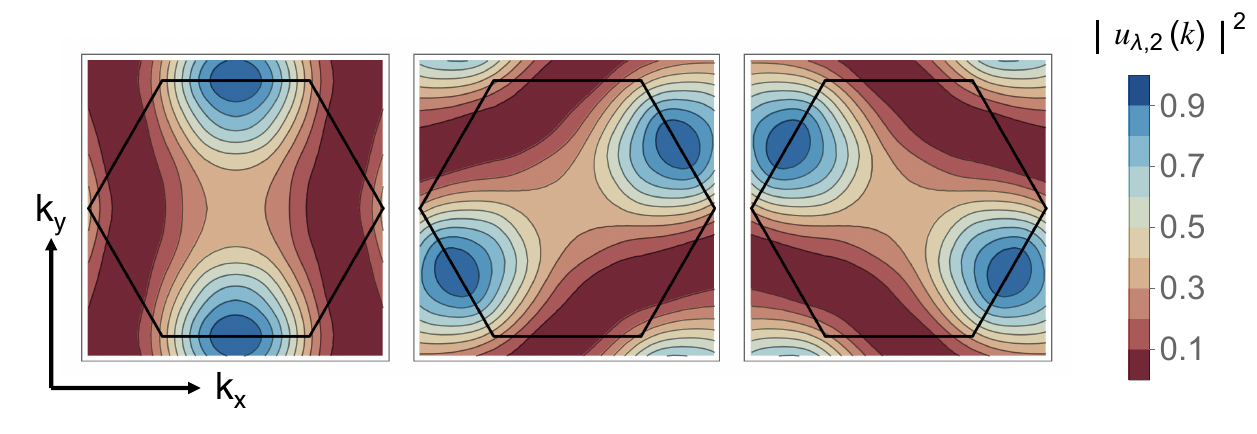}
    \caption{Contour plot of the sublattice weights for the second band corresponding to the sites A (left), B (middle), and C (right) of the kagome lattice at $(t_2\, \phi)$ = $(0.258,\, \pi/2 )$, demonstrating that each $\bs{M_i}$ point on band $2$ corresponds to one of the three sites of the kagome lattice. In this example, the band carries a Chern number of $+2$ while supporting HOVHS at the $\bs{M}$ points. Here, the black lines denote the first Brillouin zone boundaries.}
    \label{fig: SI}
\end{figure}

Similarly, for the other two high-symmetry points $\bs{M_2}$ and $\bs{M_3}$, the eigenvalues of the corresponding Hamiltonian yield the eigenvalues $\epsilon_{1} \leq \epsilon_{2} = 0 \leq \epsilon_{3}$, where the $\epsilon = 0$ case again corresponds to the second band. 
Owing to $\mathcal{C}_3$ rotation symmetry, the eigenvectors of the second band at $\bs{M_2}$ and $\bs{M_3}$, $\Psi_{2,\bs{M_2}}=(0 \: 0 \: 1)^T$ and $\Psi_{2,\bs{M_3}}=(1 \: 0 \: 0)^T$, are maximally localized on the C and A sublattices, respectively.
Fig.~\ref{fig: SI} shows the strong sublattice localization at each of $\bs{M_i}$ 
on the second band.

SI plays an important role in constraining the
interactions between electronic states located near $\bs{M_i}$ when the Fermi level crosses a Van Hove singularity. Earlier works have explored this regime for logarithmic VHS in kagome lattice \cite{kiesel2012sublattice,wu2023sublattice,kiesel2013unconventional}. The presence of the complex hopping on this kagome lattice, remarkably, uncovers a new regime where the second band supports HOVHS on the $\bs{M_i}$, displayed in the green contours of Fig.~\ref{fig: HOVHS Phase Diagram+Intersection}. Furthermore, these band support higher Chern numbers, $C = \pm 2, \pm 4$ (middle panel of Fig.~\ref{fig: Chern number}), generalizing the 
SI in time-reversal broken Chern bands beyond the Chern number $C = \pm 1$ band on the honeycomb lattice \cite{castro2023emergence}. The identification of topological bands showing SI and HOVHS is a promising platform to explore electronic correlations.

\section{Discussion}

Motivated by recent interest in band structures supporting higher-order Van Hove singularities, we have investigated a kagome system characterized by NN and NNN hopping, which, respectively, preserve and break time-reversal symmetry (Fig.~\ref{fig: lattice model+band structure}). The latter, similarly, to a chiral hopping on the Haldane lattice, leads to the formation of topological bands supporting Chern numbers $C = \pm 1, \pm 2, \pm 3, \pm 4$, as depicted in Fig.~\ref{fig: Chern number}.

More notably, we have performed a comprehensive analytical and numerical analysis of the band dispersions near the six high-symmetry points $\bs{\Gamma}$, $\pm \bs{K}$, and $\bs{M_i}$ ($i=1,2,3$), which uncovered a complex and diverse domain of HOVHS controlled by the magnitude and phase of the NNN hopping 
(Figs.~\ref{fig: Band structure}-\ref{fig: HOVHS Phase Diagram+Intersection}).
As such, our analysis unveils Chern bands with strong power law divergence in the DOS, $\rho(\epsilon) \sim |\epsilon|^{-\nu}$ characterized by the exponents $\nu = 1/2, 1/3, 1/4$. Such strong singularities in the density of states can imprint characteristic features in the low temperature intrinsic anomalous Hall response, when the Fermi level crosses the HOVHS (Fig.~\ref{fig: SigmaAH}).

We have explored distinct features of the kagome system worth mentioning. First, while it takes a critical value of the NNN hopping strength for HOVHS to emerge in bands $1$ and $2$, HOVHS can emerge at the $\pm \bs{K}$ and $\bs{M_i}$ out of the third (flat) band for infinitesimal strength of the NNN, as shown in Fig.~\ref{fig: HOVHS Phase Diagram+Intersection}. 
Furthermore, the NNN hopping provides a mechanism for the realization of high Chern number bands manifesting a sublattice interference whereupon electronic states of the second band are maximally localized on the A, B, and C sublattices as the momentum approaches $\bs{M_3}$, $\bs{M_1}$, and $\bs{M_2}$, respectively (Fig.~\ref{fig: SI}). 

The classification of HOVHS in kagome systems opens promising directions for future investigation. In particular, the interplay between band topology and large density of states provides a guiding principle to exploring correlation effects in kagome lattices, which could serve as a mechanism to realize unconventional electronic orders such as chiral topological superconductivity and fractional topological states. 
{Another interesting direction to explore would be incorporating a spin degree of freedom into our model, which can be experimentally realized in kagome FeGe metals \cite{teng2023magnetism, teng2022discovery,yin2022discovery}.}

\section*{Acknowledgments}
We thank Tianhong Lu for very useful discussions. 
This work is supported by Department of Energy, Basic Energy Science, under Award DE-SC0023327. L. P. acknowledges funding from the Women in Natural Science Fellowship of Emory University.  
\vspace{5mm}
\appendix
\section{Analytical Expressions of the HOVHS lines} \label{appendix: Perturbation Theory}

On the line $\phi=\pi$ in the $t_2$-$\phi$ parameter space, bands 2 and 3 are touching at $\bs{\Gamma}$ point, and thus need the degenerate perturbation theory (PT) treatment. We can set up the matrix
\begin{equation}
    V=
    \begin{pmatrix}
        V_{22} & V_{23}\\
        V_{23}^* & V_{33}
    \end{pmatrix},
\end{equation}
where $V_{ij}=\Psi_{i,\bs{\Gamma}}^{(0)*}\hat{\mathcal{H}}(\bs{p})\Psi_{j,\bs{\Gamma}}^{(0)}$. After diagonalization, we use the two eigenvalues, $3t_2/4$ and $-1/4$, in replacement of $\Psi_{n,\bs{\Gamma}}^{(0)*}\hat{\mathcal{H}}(\bs{p})\Psi_{n,\bs{\Gamma}}^{(0)}$ for the energy shift in Eq.~\eqref{eq: PT energy}.

Given that the perturbation matrices for $\bs{M}$ and $\bs{K}$ have both linear and quadratic dependence on $\bs{p}$, we use second order perturbation theory, with the energy shift defined as the following:
\begin{equation}
    \epsilon_{n,\bs{k}+\bs{p}} - \epsilon_{n,\bs{k}}^{(0)} =  \hat{\mathcal{H}}_{nn}(\bs{p}) + 
    \sum_{m\neq n}\frac{|\hat{\mathcal{H}}_{nm}(\bs{p})|^2}{\epsilon_{n,\bs{k}}^{(0)}-\epsilon_{m,\bs{k}}^{(0)}},
    \label{eq: second order PT}
\end{equation}
where $\bs{k}$ can be $\bs{K}$ or $\bs{M}$, and 
$\hat{\mathcal{H}}_{nm} \equiv \Psi_{n,\bs{k}}^{(0)*}\hat{\mathcal{H}}(\bs{p})\Psi_{m,\bs{k}}^{(0)}$.
With Eq.~\eqref{eq: second order PT}, we derived the expressions for Hessian at $\bs{M}$ and $\pm \bs{K}$ points:
\begin{widetext}
    \begin{equation}
        \begin{split}
            \mathbb{H}_{1,\bs{M}}(t_2,\phi) &=\frac{1}{64 \gamma^6}\\
            & \times 3 \left(\gamma -9 t_2^4-9 \gamma  t_2^3 \cos (3 \phi )+(15 \gamma -4) t_2^2 \cos (2 \phi )+t_2 \left(-7 \gamma +22 t_2^2+2\right) \cos (\phi )-12 t_2^2+1\right)\\
            & \times \left(-\gamma +t_2^4+t_2 \left(-\left(\gamma +6 t_2^2+2\right) \cos (\phi )+\gamma  t_2 (t_2 \cos (3 \phi )+\cos (2 \phi ))+8 t_2 \cos ^2(\phi )\right)-1\right)
        \end{split}
    \end{equation}
    \begin{equation}
        \begin{split}
            \mathbb{H}_{2,\bs{M}}(t_2,\phi) &=\frac{1}{16 \gamma^4}\\
            & \times \left(9 t_2^3 \cos (3 \phi )-15 t_2^2 \cos (2 \phi )+7 t_2 \cos (\phi )-1\right) \left(-3 t_2^2 (t_2 \cos (3 \phi )+\cos (2 \phi ))+3 t_2 \cos (\phi )+3\right)
        \end{split}
    \end{equation}
    \begin{equation}
        \begin{split}
        \mathbb{H}_{3,\bs{M}}(t_2,\phi) &= \frac{1}{64 \gamma^6}\\
        & \times \left(9 t_2^4+\gamma  \left(1-9 t_2^3 \cos (3 \phi )\right)+(15 \gamma +4) t_2^2 \cos (2 \phi )-t_2 \left(7 \gamma +22 t_2^2+2\right) \cos (\phi )+12 t_2^2-1\right) \\
        & \times \left(3 t_2 \left(\left(-\gamma +6 t_2^2+2\right) \cos (\phi )+\gamma  t_2 (t_2 \cos (3 \phi )+\cos (2 \phi ))-8 t_2 \cos ^2(\phi )\right)-3 \left(\gamma +t_2^4-1\right)\right)
        \end{split}
    \end{equation}
    \begin{equation}
        \begin{split}
            \mathbb{H}_{1,\bs{\pm K}}(t_2,\phi) &= \frac{1}{16 t_2 \left(-6 t_2 \cot (\phi )+2 \sqrt{3} t_2+3 \csc (\phi )\right)}\csc (\phi ) \\
            & \times \left(36 t_2^3-24 t_2^2 \cos (\phi )+\sqrt{3} \left(4 t_2^2 \sin (\phi ) (5-12 t_2 \cos (\phi ))+6 t_2 \cot (\phi )-3 \csc (\phi )\right)-6 t_2\right)
        \end{split}
        \label{eq: Hessian 1K}
    \end{equation}
    \begin{equation}
        \begin{split}
            \mathbb{H}_{2,\bs{\pm K}}(t_2,\phi) &= \frac{1}{16 t_2 \left(6 t_2 \cot (\phi )+2 \sqrt{3} t_2-3 \csc (\phi )\right)}  \csc(\phi )\\
            & \times \left(-36 t_2^3+24 t_2^2 \cos (\phi )+\sqrt{3} \left(4 t_2^2 \sin (\phi ) (5-12 t_2 \cos (\phi ))+6 t_2 \cot (\phi )-3 \csc (\phi )\right)+6 t_2\right)
        \end{split}
        \label{eq: Hessian 2K}
    \end{equation}
    \begin{equation}
        \begin{split}
            \mathbb{H}_{3,\bs{\pm K}}(t_2,\phi) &= \frac{1}{8} \left(\frac{3-6 t_2 \cos (\phi )}{8 t_2^2 \cos (2 \phi )+4 t_2^2-12 t_2 \cos (\phi )+3}+6 t_2 \cos (\phi )-1\right)
        \end{split},
        \label{eq: Hessian 3K}
    \end{equation}
\end{widetext}
where $\gamma = \sqrt{t_2^2-2 t_2 \cos (\phi )+1}$.
At specific $(t_2,\phi)$ values in the parameter space, the $\bs{\pm K}$ points on either pair of neighboring bands will touch. Due to the no-level crossing theorem under the PT framework, we want to emphasize the $\mathbb{H}_{n,\bs{K}}$ expressions are only valid in some regions of the parameter space. The band-crossing happens at a set of $(t_2,\phi)$ values, related by the function $t_2 = \frac{3}{2 \left(\sqrt{3} \sin (\phi)+3 \cos (\phi )\right)}$. Starting from the NN kagome model, i.e., $(t_2, \phi)=(0,0)$, once the critical $(t_2,\phi)$ line is crossed, the Hessian expressions for $\bs{\pm K}$, as given in Eqs.\eqref{eq: Hessian 1K}-\eqref{eq: Hessian 3K}, become mixed and do not correspond to the correct band index $n$. However, we wish to point out that upon plotting the roots of all three $\mathbb{H}_{n,\bs{\pm K}}$ expressions, we do obtain the complete set of HOVHS lines for the $\bs{\pm K}$ points in the parameter space.

\bibliography{reference}

\end{document}